\documentclass[a4paper]{article}

\usepackage{graphicx}
\usepackage{lmodern}
\usepackage[varg]{txfonts} 
\usepackage[latin1]{inputenc}
\usepackage[affil-it]{authblk}

\makeatletter
\def\@maketitle{%
  \newpage
  \null
  \vskip 2em%
  \begin{center}%
  \let \footnote \thanks
    {\Large\bfseries \@title \par}%
    \vskip 1.5em%
    {\normalsize
      \lineskip .5em%
      \begin{tabular}[t]{c}%
        \@author
      \end{tabular}\par}%
    \vskip 1em%
    {\normalsize \@date}%
  \end{center}%
  \par
  \vskip 1.5em}
\makeatother

\title{Monte Carlo Studies of Identified Two-particle Correlations in p-p and Pb-Pb Collisions}

\author[1]{Gyula Benc\'edi\thanks{\texttt{bencedi.gyula@wigner.mta.hu}; Corresponding author}}
\author[1]{Gergely G\'abor Barnaf\"oldi\thanks{\texttt{barnafoldi.gergely@wigner.mta.hu}}}
\author[2]{Levente Moln\'ar\thanks{\texttt{Levente.Molnar@cern.ch}}}
\affil[1]{Wigner Research Centre for Physics of the Hungarian Academy of Sciences, Budapest, Hungary}
\affil[2]{Institut Pluridisciplinaire Hubert Curien, Strasbourg, France}

\date{Dated: \today}

\begin{document}

\maketitle

\begin{abstract}
Azimuthal particle correlations have been extensively studied in the past at various collider energies in p-p, p-A, and A-A collisions. Hadron-correlation measurements in heavy-ion collisions have mainly focused on studies of collective (flow) effects at low-$p_T$ and parton energy loss via jet quenching in the high-$p_T$ regime. This was usually done without event-by-event particle identification. In this paper, we present two-particle correlations with identified trigger hadrons and identified associated hadrons at mid-rapidity in Monte Carlo generated events. The primary purpose of this study was to investigate the effect of quantum number conservation and the flavour balance during parton fragmentation and hadronization. The simulated p-p events were generated with PYTHIA 6.4 with the Perugia-0 tune at $\sqrt{s}=7$~TeV. HIJING was used to generate $0-10\%$ central Pb-Pb events at $\sqrt{s_{\rm NN}}=2.76$~TeV. We found that the extracted identified associated hadron spectra for charged pion, kaon, and proton show identified trigger-hadron dependent splitting. Moreover, the identified trigger-hadron dependent correlation functions vary in different $p_T$ bins, which may show the presence of collective/nuclear effects.
\end{abstract}

\section{Introduction}
\label{sec:1}

Charged hadron-hadron correlations have been extensively studied since the ISR and S$p\bar{p}$S experiments in the mid '70's~\cite{Albrow:1976pq}. These measurements have provided a good basis for investigating fragmentation processes. Well before jet reconstruction methods were born, these correlation measurements have been applied more recently in relativistic heavy-ion collisions where full jet reconstruction is challenging, especially at low jet energies. The focus of these studies was to investigate medium effects such as back-to-back jet suppression and modifications of the fragmentation process. Additionally, the discovery of long-range correlation structures in p-p, p-Pb and Pb-Pb collisions at the LHC was based on two-particle correlation measurements~\cite{CorrCMS:2012,CorrALICE:2012,Aad:2013xma}. Various hadronization and hydrodynamical models of the space-time evolution of the hot and dense color medium, which have been employed to describe these phenomena, rely heavily on assumed parton and hadron distributions. Nevertheless, the measured hadron spectra and extracted fragmentation functions are integrated distributions which mix up all possible contributions of parton-hadron channels via the convolution of initial-state, QCD and final state effects of high energy collisions~\cite{Ellis:2003}. The only way to decouple the specific hadronization contributions is to test the conservation of quantum numbers by tracing the baryon number and flavor-specific emission patterns of hadrons from the partonic medium.

This study focuses on the identified-hadron correlations, which is related to the hadronization mechanisms. Here, the baryon-to-meson ratio (for light and strange particles) is enhanced by more than a factor of three in Pb-Pb collisions compared to p-p collisions~\cite{corr:1}. Since this unique kinematic range separates the low momentum thermal bulk production from the parton fragmentation dominated regime at high momenta, the expected production mechanism is of particular interest. On the other hand, the question of quantum number conservation during the phase transition from the partonic to the hadronic world can be addressed only through correlations beyond the simple baryon-to-meson ratios.

In this short review we investigate the strength of the expected effects revealed by PID-triggered hadron-hadron correlations in high-energy p-p and Pb-Pb collisions at LHC energies in the mid-rapidity region ($|\eta|<1$). This region represents the sensitive geometrical areas of the RHIC/LHC detectors, that are common to STAR, PHENIX, ALICE, ATLAS and CMS.

\section{Monte Carlo simulations of identified hadron-hadron azimuthal correlations in p-p and Pb-Pb collisions}
\label{sec:2}

\subsection{Monte Carlo samples used for the analysis}
\label{subsec:0}

In this study we focused on the generated identified particles: $\pi^{\pm}$, $K^{\pm}$, $p$, $\bar{p}$, and the charged hadrons $(h^\pm)$ as a reference.

Event generators PYTHIA~6.4~\cite{Sjostrand} and HIJING~1.36~\cite{GyulassyHijing} were used for the production of p-p and Pb-Pb event samples. The Perugia-0 tune~\cite{Skands} was chosen for PYTHIA. HIJING event generation included the simulated effects of quenching and shadowing. The generated data include 200 million p-p events at $7$~TeV and 4 million central ($0-10\%$) Pb-Pb events at $2.76$ $A$TeV.
All charged final-state particles were kept for analysis from the Monte Carlo events, nevertheless the trigger particles and associated particles used for the two-particle correlations measurements were limited in pseudorapidity. Trigger particles with $|\eta_{trig}|>0.5$ and the associated particles with $|\eta_{assoc}|>1$ were rejected.
In order to exclude the low-momentum region from the bulk effects, the transverse momenta of the trigger particles were considered above $p_{T,trig}>2$~GeV/c in all the cases. The associated particle momenta were chosen such that $p_{T,assoc} < p_{T,trig}$ to avoid double counting. 
The uncorrelated background was subtracted by the ZYAM method~\cite{zyam}.

\subsection{The definition of the associated per trigger yield}
\label{subsec:0b}

Measuring quantum number conservation requires constraints both on the trigger and the associated particle directions to tighten the sensitivity of the measurement. The associated per trigger yield is defined with the following quantity as a function of pseudorapidity difference $\Delta\eta= \eta_{trig}-\eta_{assoc}$ and azimuthal angle difference $\Delta\phi = \phi_{trig}-\phi_{assoc}$:
\begin{equation}
\label{eq:assyield}
\frac{\mathrm{d^{2}}N}{\mathrm{d}(\Delta\eta)\mathrm{d}(\Delta\phi)}(\Delta\phi,\Delta\eta,p_{T,trig},p_{T,assoc}) = \frac{1}{N_{\mathrm{trig}}}\cdot\frac{\mathrm{d}N_{\mathrm{assoc}}}{\mathrm{d}(\Delta\eta)\mathrm{d}(\Delta\phi)}~,
\end{equation}
where $N_{\mathrm{trig}}$ and $N_{\mathrm{assoc}}$ is the number of trigger and associated particles, respectively. This quantity was measured in several $p_{T,trig}$  and $p_{T,assoc}$ intervals. The azimuthal angle correlations in this study are always projected within the pseudorapidity difference $|\Delta\eta|<1$. The associated per trigger yield on the near-side was extracted from the $\Delta\phi$ interval $|\Delta\phi|<\pi/2$ and the away-side from the $\Delta\phi$ interval $\pi/2 < \Delta\phi < 3\pi/2$.

\subsection{The PID-associated spectra}
\label{subsec:1}

\begin{figure}[!htb]
		\centering
		{\includegraphics[width=1.0\linewidth]{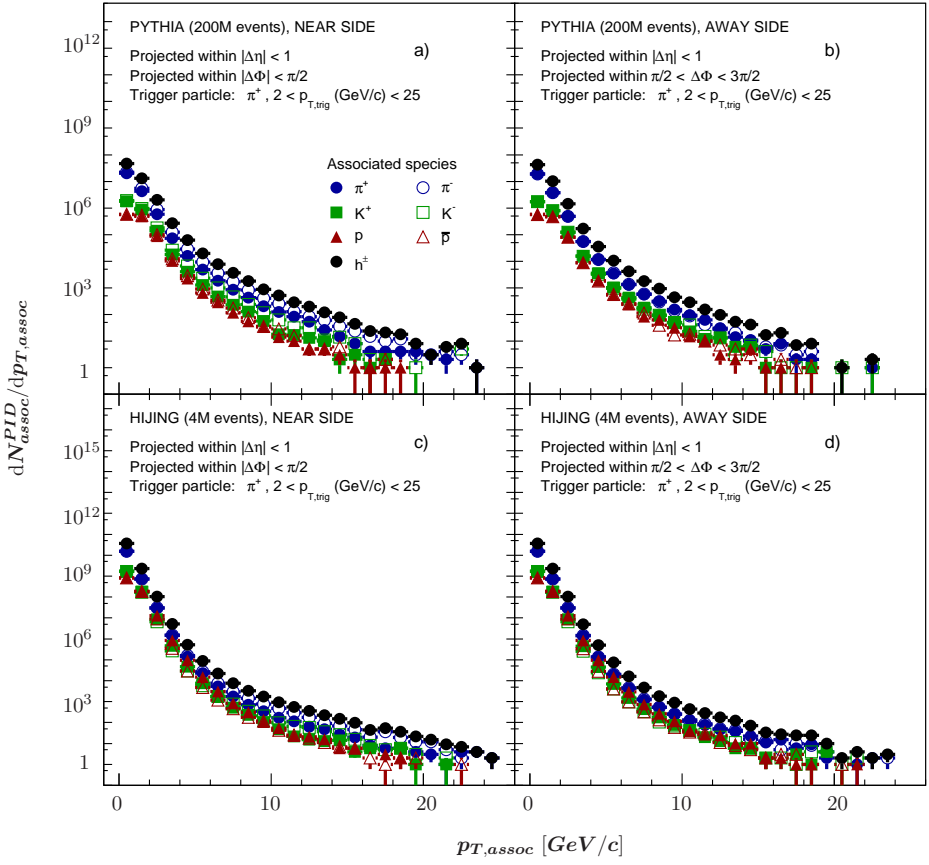}\label{fig:1}}
	    	\caption{(Color online) Identified associated particle spectra in correlation with positively charged pions ($\pi^{+}$). Open symbols indicate the positive particles and the filled symbols indicate the negative particles. The upper row ($a$ and $b$) shows the PYTHIA simulation in p-p and the lower row ($c$ and $d$) shows the $0-10\%$ central Pb-Pb HIJING events. The left column corresponds to the near side ($\pi/2 < |\Delta\phi|$) correlations and the right column corresponds to the away-side ($\pi/2 < \Delta\phi < 3\pi/2$) correlations.
		}
		\label{fig:PID-triggered_spectra}
	\end{figure}

To test the quantum number conservation with $p_{T,assoc}$ the identified associated spectra of the associated particles (hereafter called PID-associated spectra) have been plotted. We have investigated the PID-associated raw spectra using $\pi^{\pm}$, $K^{\pm}$, $p$, $\bar{p}$, and charged hadron $h^{\pm}$ triggers. Figure~\ref{fig:PID-triggered_spectra} shows the raw (unnormalized) identified particle yields $\mathrm{d} N^{PID}_{assoc}/ \mathrm{d} p_{T,assoc}$ for the identified associated particles up to $p_{T,assoc}<25$ GeV/c.
The PID-associated spectra have been plotted for positively charged pions, as trigger particles, selected from the transverse momentum range \mbox{$2$ GeV/c $ < p_{T,trig} < 25$ GeV/c}. The upper row (Figure~\ref{fig:PID-triggered_spectra} $a$ and $b$) shows the $7$ TeV p-p PYTHIA results and the lower row (Figure~\ref{fig:PID-triggered_spectra} $c$ and $d$) shows the $0-10\%$ central Pb-Pb collisions at 2.76 $A$TeV from HIJING. The left column shows the spectra extracted from the near-side and the right column shows the spectra extracted from the away-side.
The PID-associated spectra of the different associated particle species show the expected evolution with the transverse momentum. The yields of the PID-associated spectra significantly decrease with the selection of charged pion, kaon and proton triggers.

\subsection{Identified particle ratios}
\label{subsec:2}

At the Monte Carlo event generator level, the basic conservation laws of the quantum numbers\,--\,such as charge ($C$), baryon number ($B$), and strangeness ($S$)\,--\,are fulfilled and reflected in the PID-associated spectra after the hadronization process. To extract and enhance the expected quantum number conservation effects, we use the following ratio:
\begin{equation} 
\frac{\mathrm{d}N^{PID}_{assoc}}{\mathrm{d}p_{T,assoc}}~\Big/~\frac{\mathrm{d}N^{hadron}_{assoc}}{\mathrm{d}p_{T,assoc}}~
\end{equation}
which is the PID-triggered to charged hadron-triggered ratio of the PID-associated particle spectra, that is the case when we plot the associated spectra having an identified trigger compared to the one when we have a unidentified (charged hadron) trigger. Furthermore, it can be understood as a specific normalization to the hadron-triggered associated spectra.\\

	\begin{figure}[!htb]
		\centering
		{\includegraphics[bb = 0 0 477 454, clip, width=1.0\linewidth]{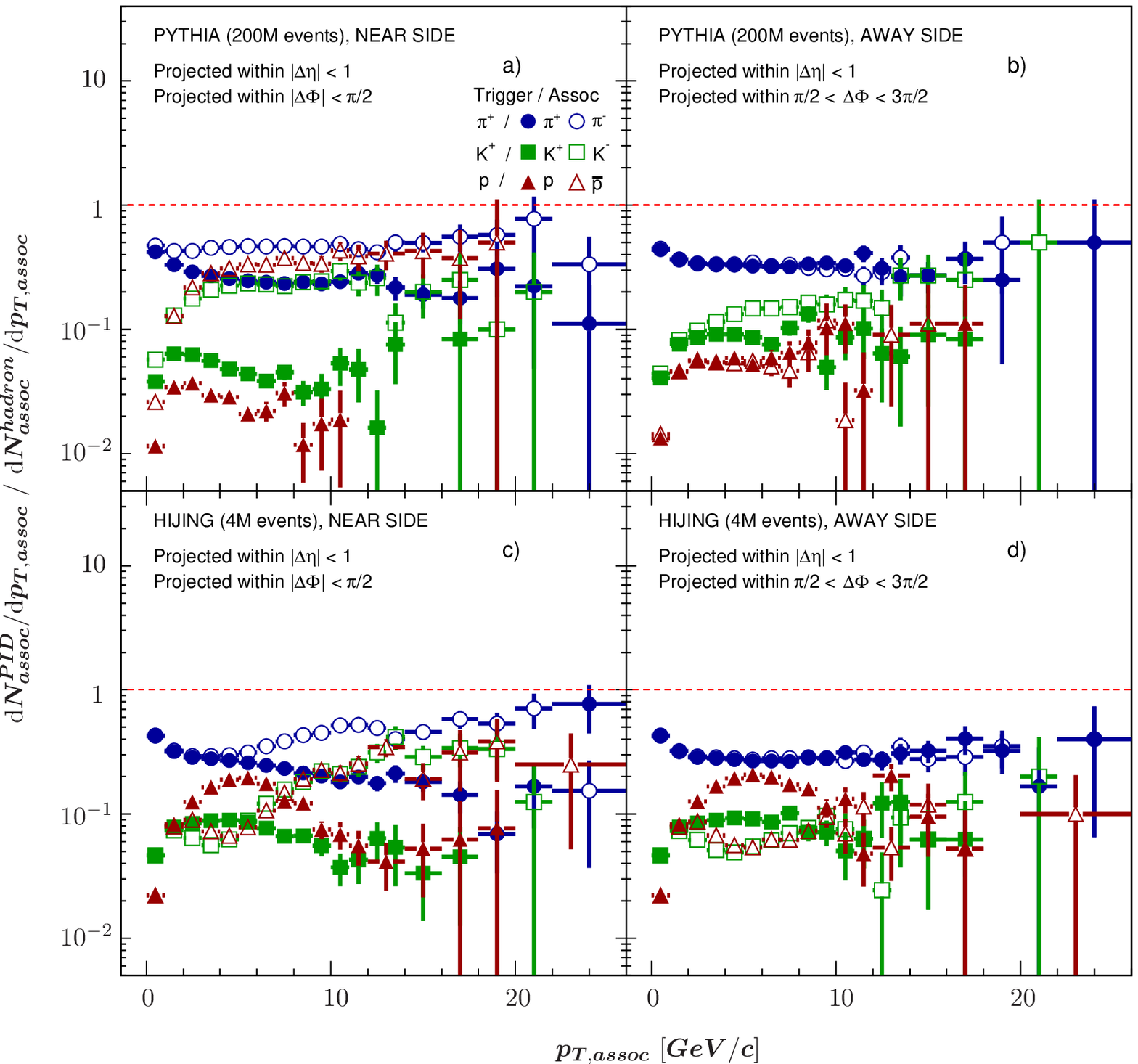}}
		\caption{(Color online) PID-triggered associated particle yields relative to the charged hadron-triggered associated yields. Open symbols indicate the positive particles and the filled symbols indicate the negative particles. The three different cases for the trigger/associated particle species are the unlike-sign pairs of $\pi^{+}/\pi^{-}$, $K^{+}/K^{-}$ and $p/\bar{p}$. The upper row ($a$ and $b$) shows the PYTHIA simulation in p-p and the lower row ($c$ and $d$) shows the $0-10\%$ central Pb-Pb HIJING events. The left column corresponds to the near side ($\pi/2 < |\Delta\phi|$) correlations and the right column corresponds to the away-side ($\pi/2 < \Delta\phi < 3\pi/2$) correlations.}
\label{fig:PID-triggered_ratio}
	\end{figure}

In Figure~\ref{fig:PID-triggered_ratio} the ratios for six identified trigger~/~identified associated particle species are shown: $\pi^{+}/\pi^{+}$, $\pi^{+}/\pi^{-}$, $K^{+}/K^{+}$, $K^{+}/K^{-}$, $p/p$, and $p/\bar{p}$. The Fig.~\ref{fig:PID-triggered_ratio}~$a$ shows the near-side ratios in p-p events. A visible splitting effect is observed for all the trigger species starting from the first $p_{T,assoc}$ bin up to $10-12$ GeV/c. The largest splitting is observed in the unlike-sign correlation, $p/\bar{p}$, where there is approximately $10$-times more anti-protons than protons above a specific $p_{T,assoc}$ bin, that is the splitting strength is about an order of magnitude larger than the like-sign correlation $p/p$. The splitting, in case of unlike-sign pairs, for $K^{+}/K^{-}$ is weaker than in the $p/\bar{p}$ but still comparable with it but the $\pi^{+}/\pi^{-}$ shows very weak splitting.

In contrast, the away-side p-p ratios, in Fig.~\ref{fig:PID-triggered_ratio}~$b$, show an equal ratio for the pions and protons (compared to the near-side in Fig.~\ref{fig:PID-triggered_ratio}~$a$). Conversely, kaons show a weak splitting in a limited transverse momentum range with respect to the near-side.
The baryon number and charge is conserved and leads to highly correlated distributions in the same cone, since the fragmentation process was not modified by the media. These qualitative observations show that the strength of the quantum number conservation effects is increasing in the order of charge ($C$), strangeness ($S$) and baryon number ($B$) respectively.
In case of hadron correlations with multiple quantum number conservation ($C$, $B$, $S$, \dots~etc.) this effect will be stronger.

Fig.~\ref{fig:PID-triggered_ratio}~$c$~and~$d$ exhibit a peculiar pattern for the case of HIJING on the near- and away-side in the $p_{T,assoc}=2-8$ GeV/c range for charged kaons and protons, where the baryon/meson anomaly was observed at RHIC and at the LHC~\cite{ALICEbaryon-meson}, followed by the reversal of the splitting trend at higher transverse momenta. The charged pions show the splitting pattern only on the near-side.

The lack of the splitting effect below $2$ GeV/c may be attributed to the fact that in HIJING the embedded mini-jet production starts to appear around this scale. On the away-side (Fig.~\ref{fig:PID-triggered_ratio} $d$), it can be seen that the effect is opposite to the PYTHIA (Fig.~\ref{fig:PID-triggered_ratio} $b$), there are more like-sign particles coming from that region. Although it is much smaller than in the near-side, but not negligible.
Moreover, we believe that the observed splitting of the protons and anti-protons presumably related to the embedded parametrization of baryon number and charge in HIJING. Further studies are ongoing to characterize better the observed splitting effects that provide an interesting contrast point between various MC event generators and experimental results from the LHC.

\subsection{Identified particle ratios in different trigger $p_T$ bins}
\label{subsec:3}

In the previous subsection the associated particle spectra were studied in the case that the trigger particles have $p_{T}$ in a wide range (\mbox{$2$ GeV/c $ < p_{T,trig} < 25$ GeV/c}). In this section we investigate the $p_{T,trig}$ dependence of the spectra. We plot the differences between the like-sign and unlike-sign particle pairs for the normalized yields $(\mathrm{d}N^{PID}_{assoc}/\mathrm{d}p_{T,assoc})/(\mathrm{d}N^{hadron}_{assoc}/\mathrm{d}p_{T,assoc})$ as plotted in Fig.~\ref{fig:PID-triggered_ratio}. Since positively charged triggers were choosen we came to the definition below in order to plot these ratios higher than one:

	\begin{figure}[!htb]
		\centering
		{\includegraphics[bb = 0 0 477 454, clip, width=1.0\linewidth]{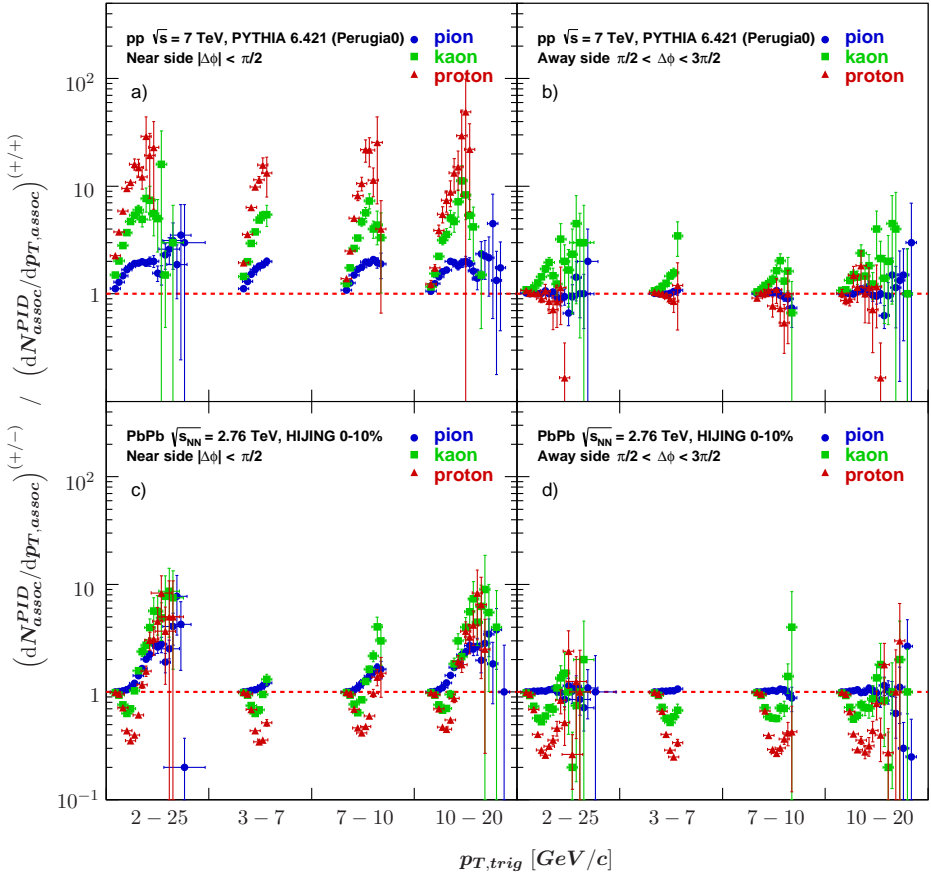}}
		\caption{(Color online) Yield differences between the oppositely charged trigger/associated particle pairs. The yield enhancement is relative to the positively charged trigger particles. The upper row ($a$ and $b$) shows the PYTHIA simulation in p-p and the lower row ($c$ and $d$) shows the $0-10\%$ central Pb-Pb HIJING events. The left and right panels correspond to the near- and away-side cases, respectively.}
	\label{fig:PID-triggered_ratio_vs_triggerPtBins}
	\end{figure}

\begin{equation}\label{eq:2}
\scalebox{1.75}{$\frac{\left(\frac{\mathrm{d}N^{PID}_{assoc}}{\mathrm{d}p_{T,assoc}}~\Big/~\frac{\mathrm{d}N^{hadron}_{assoc}}{\mathrm{d}p_{T,assoc}}\right)^{(+/-)}}{\left(\frac{\mathrm{d}N^{PID}_{assoc}}{\mathrm{d}p_{T,assoc}}~\Big/~\frac{\mathrm{d}N^{hadron}_{assoc}}{\mathrm{d}p_{T,assoc}}\right)^{(+/+)}} = \frac{\left(\frac{\mathrm{d}N^{PID}_{assoc}}{\mathrm{d}p_{T,assoc}}\right)^{(+/-)}}{\left(\frac{\mathrm{d}N^{PID}_{assoc}}{\mathrm{d}p_{T,assoc}}\right)^{(+/+)}}$}
\end{equation}

where $(+/-)$ and $(+/+)$ denote the unlike-sign and like-sign trigger/associated particle pairs, respectively. This ratio gives a more detailed insight into the evolution of the particle yield differences between the positive and negative associated particles as a function of $p_{T,assoc}$. The ratios defined by Eq.~(\ref{eq:2}) are plotted for four different $p_{T,trig}$ trigger bins on the horizontal axis of Fig.~\ref{fig:PID-triggered_ratio_vs_triggerPtBins}. In Fig.~\ref{fig:PID-triggered_ratio_vs_triggerPtBins} $a$ it can be observed how the splits\,--\,in agreement with Fig.~\ref{fig:PID-triggered_ratio}\,--\,evolve as a function of $p_{T,assoc}$ in the near-side cone in PYTHIA events. In general, the splitting effect is the largest and growing for the case when we look at the differences between the $p$ and $\bar{p}$. Concerning strange particles it seems to saturate at higher $p_{T,assoc}$. For pions the relative difference is about factor of $2-3$ at most. There is almost no change in the ordering of the species in the other $p_{T,trig}$ bins, $3-7$~GeV/c, $7-10$~GeV/c, and $10-20$~GeV/c, and the shapes remain at the same level.

Concerning the away-side, in Fig.~\ref{fig:PID-triggered_ratio_vs_triggerPtBins}~$b$, pions and protons with opposite charges are generated with the same yield but there can be a marginal difference for the kaons as it was shown in Fig.~\ref{fig:PID-triggered_ratio}~$b$. For HIJING-generated Pb-Pb events in the $0-10\%$ central region, Fig.~\ref{fig:PID-triggered_ratio_vs_triggerPtBins} $c$ and $d$, the kaons and the protons show an reverse trend with respect to the PYTHIA events.
In the near-side case, the spectrum is different for the first $p_{T,assoc}$ bins in all cases for the trigger $p_{T,trig}$ bins and after that rises above unity. The number of $K$'s produced (compared to $\pi$'s) is proportional to $p_{T,assoc}$.

This trend is unexpected and completely opposite to the PYTHIA on the away-side for all the $p_{T,trig}$ bins. Moreover, the effect manifests with similar magnitude within any $p_{T,trig}$ ranges. In summary, these effects are observed in the central Pb-Pb collisions within the $3-10$ GeV/c transverse momentum range, where the Cronin effect~\cite{Cronin:1977} and the baryon/meson anomaly act. Conversly these effects are not seen in the bulk, as indicated by the plot.

\subsection{Testing conservation laws by $\Delta\phi$ correlations}
\label{subsec:4}

\begin{figure}[!htb]
		\centering
		{\includegraphics[bb = 114 241 580 675, clip, width=1.0\linewidth]{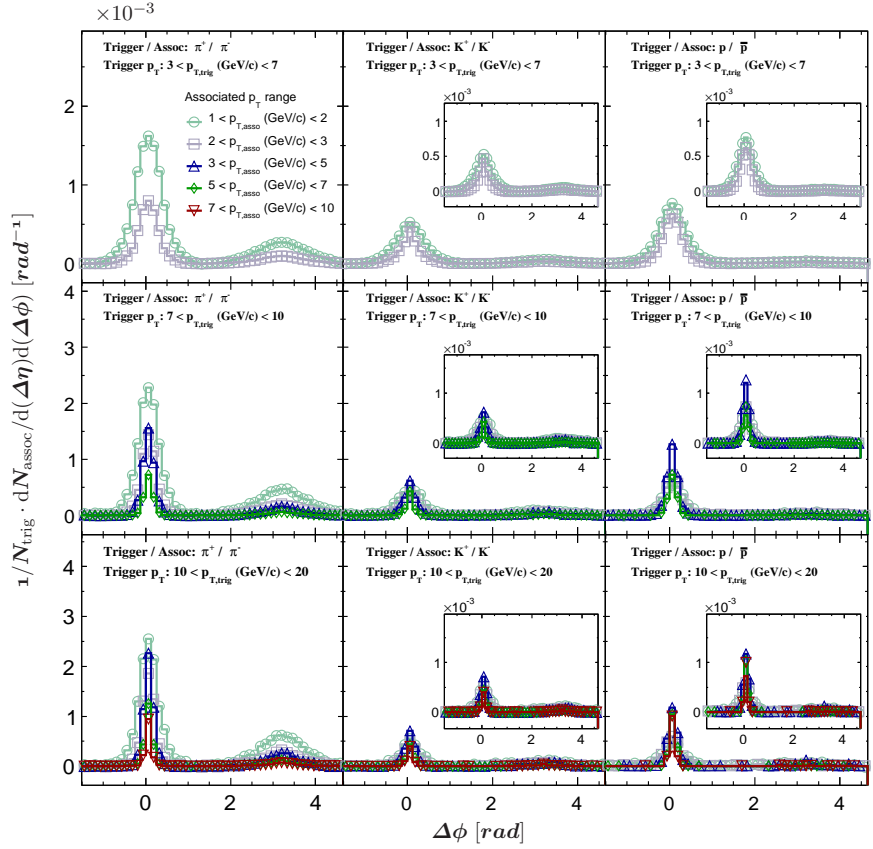}}
		\caption{(Color online) PYTHIA simulation of the p-p collisions at $\sqrt{s}=7$~TeV. The associated per trigger yields are plotted for unlike-sign trigger/associated particle pairs: $\pi^{+}/\pi^{-}$, $K^{+}/K^{-}$ and $p/\bar{p}$ in different $p_{T,trig}$\,--\,$p_{T,assoc}$ bins. The extracted distributions are integrated in the pseudorapidity difference range $|\Delta\eta|<1$. The different columns and rows correspond to different species of trigger/associated particle pairs and trigger momentum bins $p_{T,trig}$, respectively. The insets correspond to magnified portion of the y-axis of the plotted distribution for better visibility.} 
		\label{fig:phi_distr_pythia}
	\end{figure}

In Fig.~\ref{fig:phi_distr_pythia} the $\Delta\phi$-distributions given by Eq.~(\ref{eq:assyield}) of identified two-particle correlations are shown for unlike-sign trigger/associated particle pairs in different $p_{T,trig}$, $p_{T,assoc}$ bins extracted from PYTHIA events. Recall that the spectra, mentioned previously, are baseline-subtracted by applying the ZYAM-method. The columns refer to the different species of the trigger/associated particle pairs, going from left to right we plotted $\pi^{+}/\pi^{-}$, $K^{+}/K^{-}$ and $p/\bar{p}$. The rows correspond to the different settings of the $p_{T,trig}$ bins, going from top to bottom they are the following: \mbox{$3$ GeV/c $ < p_{T,trig} < 7$ GeV/c}, \mbox{$7$ GeV/c $ < p_{T,trig} < 10$ GeV/c}, and \mbox{$10$ GeV/c $ < p_{T,trig} < 20$ GeV/c}.

The near- and away-side peaks show the expected evolution of $p_{T,assoc}$. Higher trigger momentum $p_{T,trig}$ yields a narrower near-side peak, whereas higher associated momentum particles $p_{T,assoc}$ result in a lower associated particle yield.
Nevertheless, the amplitude of the correlation peak stays approximately constant on the away-side as a function of $p_{T,trig}$. If we compare the columns, that is the different trigger/associated particle pairs, we can observe that on the near-side, the ratio of the number of the associated particles to the trigger particles is less for strange particles with respect to the baryons compared to the pions.
There is a hint that the same effect is visible on the away-side, but a more thorough study is needed.

\begin{figure}[!htb]
		\centering
		{\includegraphics[bb = 114 241 580 675, clip, width=1.0\linewidth]{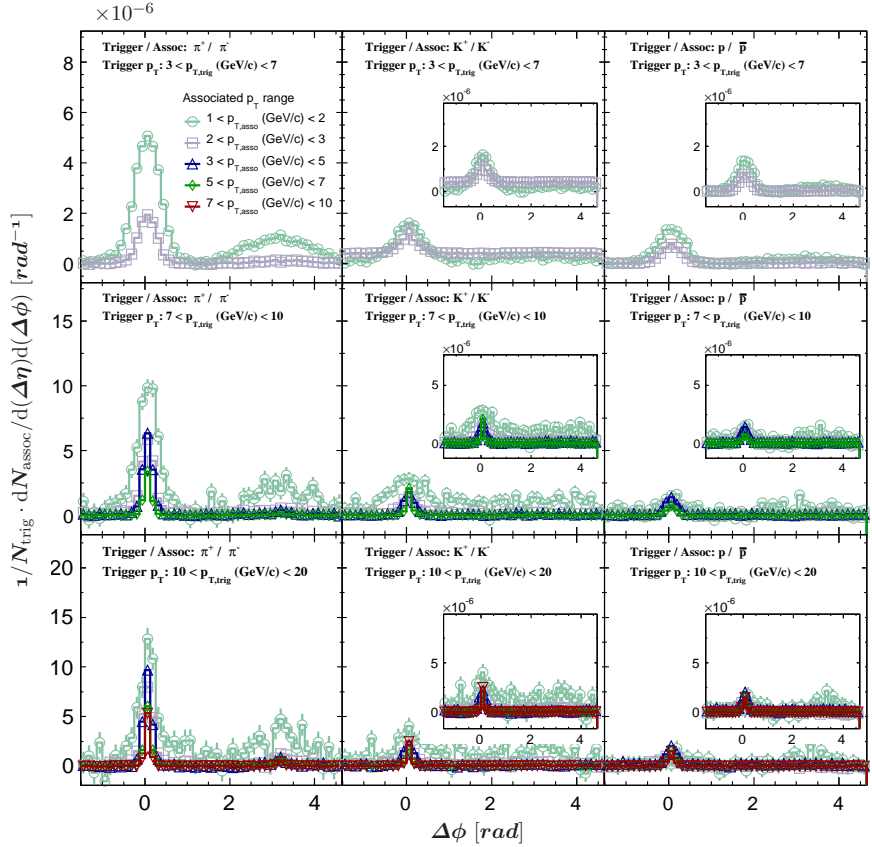}}
		\caption{(Color online) HIJING simulation of Pb-Pb collisions at $\sqrt{s_{\rm NN}}=2.76$~TeV. The associated per trigger yields are plotted for unlike-sign trigger/associated particle pairs: $\pi^{+}/\pi^{-}$, $K^{+}/K^{-}$ and $p/\bar{p}$ in different $p_{T,trig}$--$p_{T,assoc}$ bins. The extracted distributions are integrated in the pseudorapidity difference range $|\Delta\eta|<1$. The different columns and rows correspond to different species of trigger/associated particle pairs and $p_{T,trig}$ trigger momentum bins, respectively. The insets correspond to magnified portion of the y-axis of the plotted distribution for better visibility.} 
		\label{fig:phi_distr_hijing}
	\end{figure}

In Fig.~\ref{fig:phi_distr_hijing} the distributions for the Pb-Pb events, generated by HIJING, show the same trend that is seen on the near-side for the PYTHIA events.

The difference between PYTHIA and HIJING in terms of the chosen trigger/assoc-iated species is evident, practically there is only a small flavour and charge correlation effect in HIJING at all studied $p_{T,trig}$ momentum bin. Moreover, the number of associated particles for kaons and protons has the same order of magnitude on the near-side as well as on the away-side.

\section{Summary and Outlook}

Monte Carlo studies of two-particle azimuthal correlations have been performed with identified trigger particles and identified associated particles in p-p collisions at $\sqrt{s} = 7$~TeV and in Pb-Pb collisions at $\sqrt{s_{\rm NN}} = 2.76~A$TeV at mid-rapidity. As for the first verification of the generated samples the identified hadrons were extracted from the raw associated particle spectra: $\pi^{+}$, $\pi^{-}$, $K^{+}$, $K^{-}$, $p$ and $\bar{p}$ with fixed positively charged pion ($\pi^{+}$) trigger in two separated $\Delta\phi$ regions, near- and away-side. Further investigation was needed to look deeper into the validity of the quantum number conservations of correlated particle pairs. For this reason we plotted the identified associated spectra for like-sign and unlike-sign trigger/associated particle pairs compared to the charged hadron identified associated spectra. We obtained conclusive results for PYTHIA: the PID-triggered-to-hadron-triggered associated spectra on the near-side show a significant splitting between the particles and anti-particles which have the largest effect for the baryon triggers, but is absent on the away-side. Qualitatively the difference between the baryon triggers and the pions is about one order of magnitude. For the HIJING simulation the results show an opposite trend to the PYTHIA up to $p_{T,assoc} = 7-10$ GeV/c and these effects are not vanishing on the away-side except for the pions. The splitting in Pb-Pb collisions starts at slightly larger $p_{T,assoc}$ threshold than in p-p because of the embedded physics in HIJING (for example the minijet production), but further investigation is needed with real data to confirm this observation. To conclude: collective/nuclear effects are visible in the studied $3-10$~GeV/c momentum range.\\
There is no significant change between the yields of the unlike-sign particle pairs as we change the trigger $p_{T,trig}$ bins. Finally, we showed the PID-triggered azimuthal $\Delta\phi$ correlation spectra for the different trigger species. We have seen more associated protons than kaons on the near-side of the p-p collisions which does not depend on the applied trigger momentum $p_{T,trig}$ cuts. In Pb-Pb collisions the probability to generate protons or kaons as associated particles on the near-side is about equal.

\section*{Acknowledgement}

This work was supported by Hungarian OTKA grants NK778816, NK106119,
K104260, and TET 12 CN-1-2012-0016. Author Gergely G\'abor Barnaf\"oldi also thanks the J\'anos Bolyai Research Scholarship of the Hungarian Academy of Sciences.



\begin{thebibliography}{9}

\bibitem{Albrow:1976pq}
 M.~G.~Albrow {\it et al.},
  Phys.\ Lett.\ B {\bf 65} (1976) 295.

\bibitem{CorrCMS:2012}
  
  S.~Chatrchyan {\it et al.}  [CMS Collaboration],
  [arXiv:1310.8651 [nucl-ex]].
  
\bibitem{CorrALICE:2012}

 B.~Abelev {\it et al.}  [ALICE Collaboration],
  Phys.\ Lett.\ B {\bf 719} (2013) 29
  [arXiv:1212.2001 [nucl-ex]];
B.~B.~Abelev {\it et al.}  [ALICE Collaboration],
  Phys.\ Lett.\ B {\bf 726} (2013) 164
  [arXiv:1307.3237 [nucl-ex]].

\bibitem{Aad:2013xma}
  G.~Aad {\it et al.}  [ATLAS Collaboration],
  JHEP {\bf 1311} (2013) 183
  [arXiv:1305.2942 [hep-ex]].

\bibitem{Ellis:2003}
R.~K.~Ellis {\it et al.} QCD and Collider Physics, Cambridge Monographs on Particle Physics, Nuclear Physics and Cosmology, 2003, p. 452 (ISBN:
    9780521545891)

\bibitem{corr:1} 
B.~B.~Abelev {\it et al.}  [ALICE Collaboration],
  Phys.\ Rev.\ Lett.\  {\bf 111} (2013) 222301
  [arXiv:1307.5530 [nucl-ex]].

\bibitem{Sjostrand}
T.~Sjostrand {\it et al.} PYTHIA 6.4 Physics and Manual, JHEP 05 (2006) 026. arXiv:hep-ph/0603175

\bibitem{GyulassyHijing}
X.-N.~Wang, M.~Gyulassy, 
 Phys.\ Rev.\ {\bf D44} (1991), 3501\,--\,3516.

\bibitem{Skands}
P.~Z.~Skands,
Phys. Rev. {\bf D82} (2010) 074018
arXiv:hep-ph/1005.3457

\bibitem{zyam}
 A.~Adare {\it et al.}  [PHENIX Collaboration],
  Phys.\ Rev.\ C {\bf 78} (2008) 014901
  [arXiv:0801.4545 [nucl-ex]];

\bibitem{ALICEbaryon-meson}
H.~Appelshauser,
  J.\ Phys.\ G {\bf 38}, 124014 (2011)

\bibitem{Cronin:1977}
Antreasyan, D. {\it et al.}
Phys. Rev. Lett. {\bf 38} (1977)

\bibitem{VhmpidLoI:2013}
\textit{A Very High Momentum Particle Identification Detector (VHMPID) for ALICE, Letter of Intent},
	arXiv:1309.5880 [nucl-ex]

\end{thebibliography}
\end{document}